\documentclass[a4paper]{article} 
\usepackage[affil-it]{authblk}
\usepackage[english]{babel} 
\usepackage{fancyhdr} 
\usepackage{graphicx} 
\usepackage{float}
\usepackage{mhchem} 
\usepackage[margin=1.2in]{geometry}

\title{Calibration of highly segmented, compact gamma camera for Molecular Breast Imaging} 
\vspace{3cm}

\author[3]{A. Marcucci}
\author[1,2]{F. Garibaldi}
\affil[1]{Istituto Superiore di Sanit\`a, viale Regina Elena, 299 - 00161 Rome - Italy}
\affil[2]{Istituto Nazionale di Fisica Nucleare, piazzale Aldo Moro, 5 - 00185 Rome - Italy}

\author[3]{G. Limiti}
\author[1]{T. Insero}

\affil[3]{Sapienza, Unversit\`a di Roma, piazzale Aldo Moro, 5 - 00185 Rome - Italy}

\author[4]{P. Musico}
\affil[4]{Istituto Nazionale di Fisica Nucleare, via Dodecaneso, 33 - 16146 Genova - Italy}

\author[1]{E. Cisbani}

\date{30 October 2018} 

\begin{document}

\maketitle

\begin{abstract}
  Breast cancers is the second leading cause of cancer mortality in women; early diagnosis increase the probability of a 
  successful therapy; any marginal improvement in this direction helps sparing lives. 
  Imaging plays a major role in diagnosis: different morphological imaging methodologies are currently available:
  from the widely used screening 2D mammography,
  to increasingly adopted 3D tomosynthesis, from ultrasound to expensive contrast-enhanced magnetic resonance.
  None of them guarantee the desired (100\%) sensitivity and specificity and most of them fail in dense breast,
  where cancer incidence is higher.
  
  In this context functional imaging techniques such as Molecular Breast Imaging (MBI) represents an important supplemental screening,
   especially in the more questionable cases.
  Dedicated compact gamma cameras, with single or dual symmetric heads
  are commercially available, offering performances independent from the density of breast,
  larger specificity than morphological mammographic imaging, but
  relatively modest sensitivity to sub-millimeter tumors.

  In order to further extend the MBI performances an innovative asymmetric dual detector device,
  with mixed optics has been recently proposed and prototyped; the sensors are highly segmented
  with a correspondingly large number of independent, configurable,
  electronic readout channels with self-triggering capability.
  
  This flexible electronics architecture has different advantages in addition
  to those related to the adopted asymmetric dual detector geometry:
  real-time event selection based on the adjustable gain and discriminator threshold
  at single channel (or group of channels) level;
  repeatable, quick hardware and software channel response equalization;
  configurable list mode acquisition for versatile offline image processing.
  These benefits come at the expenses of more complex calibration methods and optimization procedures,
  which are detailed in the present paper.
  
\end{abstract}

\section{Introduction}
Breast cancer is the most common non-skin type malignancy and the second leading cause of cancer mortality in women
(more than 2 million new cases estimated in 2018 \cite{globascan}).
Mammography is the most widely used screening modality for breast cancer with 70-90\% sensitivity and $\sim 90$\% specificity;
it is uneffective in {\it dense} breast which corresponds to the cases
with higher cancer risk, involving around 30\% of women \cite{bib:OConnor2}.
3-D evolution of mammography, the breast tomosynthesis, offers higher cancer detection probability ($\sim 1-3$ additional cancer per 1000 women \cite{bib:oconn15,bib:freer}) respect to the latter, with improved performances on dense breast,
however has still significant probability of false positive diagnoses.
Supplemental screening tests \cite{bib:vourtsis18} are represented by (automated) ultrasound and expensive constrast-enhanced magnetic resonance imaging
(MRI) which allow detection of additional $2-4$ and $8-18$ cancers per 1000 women respectively \cite{bib:oconn15,bib:freer}
but still suffering of relevant false positive and/or uneffectiveness on dense breast cases, ultimately requiring the use of invasive and
generally expensive biopsy, and posing questions on cost/benefit optimization\cite{bib:costs}.

Besides the above radiological imaging (mammography, ultrasound, MRI),
nuclear medicine imaging techniques are playing an increasingly role in the diagnostic characterization of breast lesions \cite{bib:Kim11}.
For this reason functional specific imaging of a radiotracer with preferential uptake on breast tumor lesions has bene proposed \cite{Khalkhali}. Standard gamma cameras are not suitable for that scope.
Only tumors larger the 1 cm are commonly detected, essentially because of limited spatial resolution and large detector to lesion distances. For this reason dedicated detectors have been proposed \cite{scop,devinc}.
Single and dual head MBI devices are commercially available; in one large screening study, the specificity was 93\% for molecular breast imaging versus 91\% for mammography (p = 0.07) \cite{bib:OConnor2}.
With these techniques, the breast is positioned between two detectors that create 2D images of the breast after administration of Technetium-99 ($\ce{^{99m}Tc}$)-sestamibi \cite{bib:tc05}.\\
Key parameters for detecting small lesions are: spatial resolution, Signal to Noise Ratio (SNR) and Contrast to Noise Ratio (CNR). Energy resolution plays only a secondary additional role in imaging breast under compression.\\
In order to improve the performances of a MBI system we have designed a dual detector
system that increases significantly the capability of detecting small tumors \cite{bib:garibaldi}.
The intrinsic properties of the gamma detector have been optimized; further improvement are foreseen for its integration into a tomosynthesis device to get almost simultaneous morphological and functional multimodality information, for superior diagnosis.

The imaging system is flexible and compact and it is made up of two Detector Heads:
the first (Big Head) consists in a large, high efficiency detector, with parallel hole collimator; the second head (Small Head),
with a small $50\times 50$ $mm^2$ detector, is used as a spot compressor, getting closer to the tumor lesion.\\
A pinhole collimator is used with further advantages in terms of spatial resolution and efficiency.

A multicenter clinical trial is planned; it will be a multimodality detector in the market and
consequently in the clinical practice.
The system has thousands of independent readout channels which offer a large flexibility in terms
of image processing and optimization; however this requires a careful multi-step calibration
of the large number of correlated free parameters (e.g gains, thresholds, timing and other readout electronics settings).
The complexity of such calibration is somehow similar to the segmented calorimeters used in nuclear and particle physics.
However the novelty of the device combined to peculiarities such as the auto-triggering mode (described later) requires the development
of new, original procedures and methods, which are presented in this paper.

\section{Experimental Setup}
\begin{figure}[H]
\centering\includegraphics[scale=0.8]{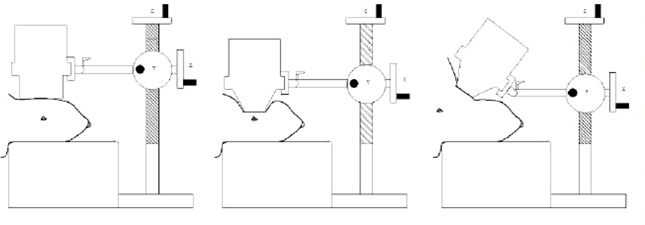} 
\caption{Prototype of the Dual Head system. The breast is positioned between the two detector heads with mild but effective compression, especially in the pin-hole geometry (middle and right images).}\label{riv}
\end{figure}

A sketch of the dual head high-resolution Gamma Camera prototype developed in our Lab for Molecular Breast Imaging and its basic modes of operations relative to a more conventional system, is shown in Fig. \ref{riv}: each of the two heads is a compact gamma camera, as described in the Introduction. The bigger detector acts as a support for the breast and has a size of a typical mammographic screen ($150 \times 200$ mm$^2$), the other is smaller ($50 \times 50$ mm$^2$), acts as a spot compressor and can be positioned in front of the region of the breast with suspected lesion, as appears in mammography but can be positioned anywhere.

The photon detector is equipped with Position Sensitive Photomultipliers (PSPMTs) Hamamatsu H8500 \cite{bib:pmt}.
The photomultipliers are made up $8 \times 8$ array of pixel. They have very high packing fraction of 89\% and a quantum effciency of 35\%. PSPMTs are powered in pairs by EMCO L12A-R High Voltage low noise DC-DC converters with negative polarity, which are in turn powered by a low voltage, up to 12V 1A, power supply. The dedicated readout electronics (detailed in the following) are able to read all channels separately ($12 \times 64 = 768$ channels in this case).
Such electronics, if properly tuned, overcomes the performances of the resistive chain readout system:
trigger is generated from the pulse of the single channel (or a logic combination of some of them),
it improves the channel to channel uniformity and a high processing capability is achieved (distortion compensation,
superior centroid estimation, background event rejection).\\
The first detector in Fig. 1 consists in a high sensitivity lead parallel hole collimator, with 19 mm length hexagonal septa; an array of pixelated NaI(Tl), 0.5 cm thick, with pixel of 1.5 mm pitch and 0.3 mm gaps, coupled to $4 \times 3$ matrix of multi-anode PMTs Hamamatsu H8500.\\
The second detector, consists in a pinhole collimator, 2 mm hole and coupled to a scintillating crystal ($50.8\times 50.8\times 6$ mm$^3$), and Hamamatsu H8500. The system can be manually adjusted to optimize the distance between the pinhole and the axis of rotation, giving the possibility to resize the camera parameters depending on measurements requirements.

\section{Readout Electronics and Data Acquisition}
In the Single Gamma Photon Counting the electronic readout acquires the charge produced on each PMT anode by the gamma photon interacting in the scintillator.
The charge amplitudes and anode positions are combined to obtain the information about the energy of the incident gamma photon and its point of interaction in the scintillator.\\
The dedicated readout electronics has been designed taking into account also compactness, high modularity and flexibility
which give the possibility to adapt the system to several application requirements, thanks to the extensive use of the FPGA (Field Programmable Gate Array).

The readout system \cite{bib:argentieri10} is organized in a hierarchical scheme with a single main board, the Controller Board (CB),
which includes a powerful dedicated FPGA (Altera Cyclone II EP2C20 series \cite{bib:altera}),
and 13 daughter boards, the Front End cards (FE), which are based on highly configurable MAROC3
(MultiAnode ReadOut Chip) ASIC by Omega (Orsay MicroElectronics Group Associated \cite{bib:maroc, bib:barillon})
that read the signals generated in the anodes of PMTs. Each FE hosts also an FPGA (Altera Cyclone II EP2C8 \cite{bib:altera}) for
MAROC3 configuration and acquisition logic, with real-time processing capability.

The MAROC3 consists of 64 independent channels and provides both the analog (charge) and binary (trigger) information for each channel.
The single channel has a low impedence pre-amplification and a configurable 8 bit gain correction
stage followed by a signal splitter over three independent lines:
\begin{itemize}
\item  a {\it binary} line characterized by a fast shaper ($\sim 20$ ns peaking time) followed by a
  discriminator with adjustable threshold;
\item an {\it analog} line with a slow shaper ($30 - 210$ ns) followed by two Sample and Hold (S$\&$H) circuits
  that permit the charge measurement through a serial multiplexer);
\item a third line which can provide up to 8 direct analog, selectable, sums:
  this line is not used in the current implementation.
\end{itemize}
Both shapers are highly configurable.

The 64 signals of the fast binary line are ORed to produce a trigger candidate for each MAROC3 card;
the internal trigger is generated if the
hit multiplicity (number of channels contributing to the candidate trigger, from all cards) is larger than a
configurable threshold (typically 3 hits).

The trigger signal is suitably delayed and it becomes the hold signal for the S$\&$H circuit which permits the serial, multiplexed, readout of the 64 analog slow channels.
The charge, which is stored on the S$\&$H circuit is then converted in 12 bit digital signal by the ADC (Analog to Digital Converter) AD~7274 from Analog Device.\\
In addition to the abovementioned internal trigger, the readout electronics may provides a pulsed trigger as well as accept external triggers that are used, mainly, for pedestal runs and dead
time estimation.

Moreover many parameters of the electronics can be configured (trigger threshold, single channel gain, slow  and fast shapers feedback components).
Different acquisition modes are available to help reducing the amount of data transferred to the acquisition node and
therefore the speed of acquisition. Such sparse readout can be obtained either dropping the ADC values that are below a configurable level (different from each channel) or
those values corresponding to channels that did not passed the internal MAROC3 discrimination.

In the full configuration the system can read 4096 channels, i.e. 64 FE cards.

\section{Data Processing}

\begin{figure}[htb]
\centering
\includegraphics[scale=0.75]{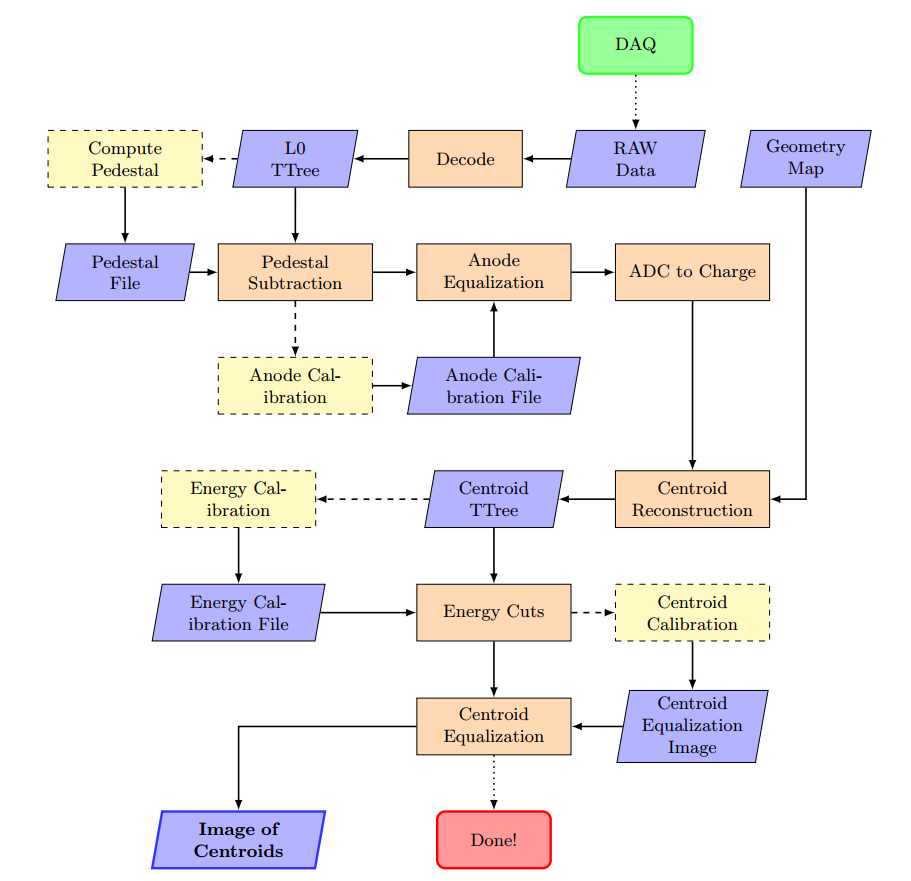} 
\caption{Flow chart of the image processor; rectangles represent processing blocks generally implemented by specific methods (dashed border distinguishes calibration dedicated processing steps); trapezoids are Input/Output data files}\label{proc}
\end{figure}

The data processing software generates the final centroid image, from the raw data, including all calibrations, equalizations and corrections that contribute to improve the final image quality.
The processor is implemented in C++ within the ROOT data analysis framework \cite{bib:root}; it
consists of two main classes that instantiate different base classes and have public methods to end up with the production of the image of centroids.
Most of the steps required for the generation of the image have intermediate outputs that represent,
under specific acquisition conditions, calibration inputs for the processing chain.
A schematic flow chart of the data processing is presented in Fig. \ref{proc}.

The data acquisition provides a raw data file in dedicated binary format, consisting of the sequence of events (list mode); each event has information on time of occurrence (with sub millisecond resolution), ADC converted charges collected on each anode channel, binary status of each channel used for the formation of the trigger. Ancillary information on the acquisition configuration and actual conditions (such as trigger rate) are stored into a companion human readable text file. 

The main processor data flow (assuming all calibrations have been computed and results stored in properly formatted files) consists of the following major steps (refer also to fig. \ref{proc}):
\begin{enumerate}
\item {\bf Raw data decoding}: the first step of data processing ingests the raw data file and the companion text file, it decodes the binary data stream into ADC values $C_i$ of each channel $i$ and then it stores them together with the other relevant information available in the input data in a structured {\verb ROOT::TTree } format, suitable for preliminary analysis and further processing.
  
\item {\bf Anode charge calibration}: the ADC values are converted into physics charges $Q_i$ by the linear expression:
    $ Q_i = g_i ( C_i - C_i^0 ) $ if $C_i > C_i^0 + n \cdot \sigma_{C_i}$ ($Q_i = 0$ otherwise),
  where the pedestal values $C_i^0$ (and channel noise $\sigma_{C_i}$) are estimated from a dedicated run\footnote{The pedestal run and its processing is pretty conventional: random or pulsed trigger is generated in the very same conditions of a normal acquisition except the sensor is not exposed to the radioactive signal source but environmental and cosmic radiation. The pedestal processor computes, for each channel, the event-average ADC value $C_i^0$ and its RMS $\sigma_{C_i^0}$ (or the related single event channel noise $\sigma_{C_i^0} \sim \sigma_{C_i}/\sqrt{N}$, with $N$ the number of events contributing to the pedestal estimation). The pedestals are estimated when either the system configuration is changed or the system is switched on. Moreover, due to the low occupancies, most of the channels in a single event are not affected by signal and
    therefore the charge spectrum of each channel in a run is similar to a pedestal spectrum and can actually be used to estimate the pedestal on an run by run basis
    (when sparse readout is not used).},
  $n=3.5$ is usually assumed,
    while the residual anode gains $g_i$ are obtained from a flood run\footnote{\label{fn:flood} In a flood run, the sensor is exposed to a spatially uniform field of radiation, which is usually obtained by a point-like radioactive source sitting at a distance of 5 or more times the linear relevant size of the sensor (as suggested by \cite{bib:iaea}).} and related residual anode equalization processor\footnote{The residual anode equalization processor computes the average of the ADC $s(\bar{C})$, pedestal subtracted spectra and then fits each anode ADC spectrum $s(\bar{C_i})$ with a simple parametrization of the average $n_i \cdot s(g_i \bar{C} - m_i)$; the estimated $g_i$ are the anode gain correction (they distribute around 1.) while the $m_i$ are checked to be within $0. \pm  3. \cdot \sigma_{C_i^0}$. The anode gains are evaluated only in case of either major system configuration changes of weekly quality control.}.
    At this level any potential common noise over multiple channels can be subtracted on an event-by-event base (pedestals can be estimated accordingly).
    
  \item {\bf Centroid reconstruction}: Once the charge of each channel (anode) is properly evaluated, the photon avalanche created by the photoelectric effect (or any other effect that release energy in the scintillator through light emission) is reconstructed by clustering the spatially contiguous anodes contributing to the avalanche. Different approaches can be considered for clustering; we adopted the ``distance from the charge maximum'', that is, the anode with maximum charge is first identified and then all the anodes whose center is at a spatial distance below a given threshold $d$ contribute to the cluster avalanche. The total charge of the cluster $C$, that is expected to be proportional to the initial energy $E_\gamma$ of the gamma photon (for optimally collected photoelastic events), is therefore given by:
    \begin{equation}
      E_\gamma \propto C = \sum_i Q_i \mbox{ with } (x_i-x_M)^2 + (y_i-y_M)^2 \le d^2
    \end{equation}
    where $x_i,y_i$ and $x_M,y_M$ are the spatial coordinate of the centers of the anode $i$ and the anode with the maximum charge respectively.
    The radius $d$ that produces good quality images, depends mainly on the scintillator geometry and PMT coupling; optimal value typically range, in our experience, between 3 and 4 times the PMT anode size.
    In addition to the gamma photon energy, the cluster centroid provides the 2-dimensional position $X_\gamma, Y_\gamma$ of the photoelectric process in the scintillator as charge weighted average of the positions of the anodes belonging to the cluster:
    \begin{equation}
      X_\gamma = \frac{ \sum _{i\in \textrm{Cluster}} Q_i x_i }{ \sum_{i\in \textrm{Cluster}} Q_i }
    \end{equation}
    
  \item {\bf Energy calibration and cuts}: the gamma energy $E_\gamma$ is obtained from the cluster total charge $C$ by the linear expression $E_\gamma = a(X_\gamma,Y_\gamma) + b(X_\gamma,Y_\gamma) \cdot C$ where $a()$ and $b()$ are calibration constants that depend on the position of the cluster centroid as described later in section \ref{sec:locenergy}. Only events whose gamma photon energy belongs to the selected photopeak range are considered for image formation.

  \item {\bf Image formation and centroid equalization}: the image representing the activity distribution of the radiopharmaceutical is finally obtained as spatial distribution of the cluster centroids of all events surviving the previous energy cuts: $I(x,y)$. This distribution is sampled on 2-dimensional bins (pixels) whose user defined size is essentially a trade-off of available statistics and desired spatial resolution. Image is ultimately corrected for local variation of the centroid distribution as discussed in section \ref{sec:cencorr}.
\end{enumerate}
It is worth to point out that the above steps (except the last, image formation) proceed on an event by event basis and therefore the introduced quantities ($C_i$, $Q_i$, $E_\gamma$, $X_\gamma$ and $Y_\gamma$) depend also on the single event considered.

\section{System Calibration}\label{sec:calib}
The independent channels readout and the highly configurable electronics offer superior flexibility for image quality optimization and application specific improvements. However the definition of the best configuration requires very carefull and complex procedures for
calibration and tuning of the many correlated parameters that are involved in the image processing.

Most of the corrections applied to the data to obtain the high quality image are related to the proper 
response of the detector to a uniform irradiation; in fact all sensitive segments (scintillator crystal pixels and PMT anodes) of the gamma imaging detector are designed to maximize homogeneity\footnote{Due to the peculiar position, the sensitive elements placed on the border of the detector active area behave sligtly differently, though they are virtually identical to the other elements.}
and one of the main goal of the calibration is their optimal equalization.

\subsection{Channels equalization}

The response of the detector depends primarily on the irradiation field, the collimator geometry, the gamma photo-conversion, position of the sensitive elements, the PMT intrinsic efficiency and high voltage, trigger threshold (both common to group of 64 channels), single channel gain, and to other electronics configuration parameters.

In the specific hardware/firmware implementation, once the collimator and the scintillator crystal have fixed, the response can be improved acting on:
\begin{itemize}
\item PMT high voltages, which affect the gain and dark rate;
  they are common to group of 64 channels;
\item electronics gains (from 0 to 4, in 256 steps) of each channel,
  and common to both the binary (trigger) and analog lines of the MAROC3; they are mainly intended to compensate the PMT anodes non-uniformity;
\item discriminator levels, which strongly affect the trigger response; they 
  apply to groups of 64 channels: one selectable threshold level for all binary lines of each MAROC3 chip.
\end{itemize}

Trigger equalization is a necessarely precondition for a proper detector response; once it is achieved, the charge uniformity can (and must) be further tuned at the software level in the image processor, by introducing the above mentioned gain correction factors.

While the PMTs are independently characterized before assembling \cite{pani}, the optimization of the thresholds and gains affecting the trigger response is a multi-variable correlated problem:
a change in a single channel gain requires the tuning of the trigger threshold which affects 64 channels and therefore a variation of their gains, which in turn may require a new threshold and additional iterations may apply.

Span the entire multidimensional phase space of thresholds ($t$) and gains ($g$) to define the optimal working point does not represent a viable solution (especially in clinics) since it would require weeks of continuous acquisition to collect the proper statistics, in stable conditions.
For this reason, a procedure has been defined that converges toward optimal $t$ and $g$ settings,
firstly shrinking the $t-g$ phase space and then scanning it in a smart way; it is made of two main steps:

\begin{enumerate}
\item estimate the optimal gains (and their upper limit) for a given threshold in single channel acquisition mode; the search is performed by binary scan of the threshold values and exploiting the linear relation between threshold and gain;
\item equalize the trigger rate of each channel for the flood (or background) irradiation constrained by the previous limits:
  in operational multihit trigger mode, perform an adaptive gain and threshold scan
  constrained by the previous findings to equalize the trigger rate of each channel for a flood (or background) irradiation

\end{enumerate}

\subsubsection{Gain-Threshold phase space shrinking}

\begin{figure}[H]
\centering
\includegraphics[scale=0.5]{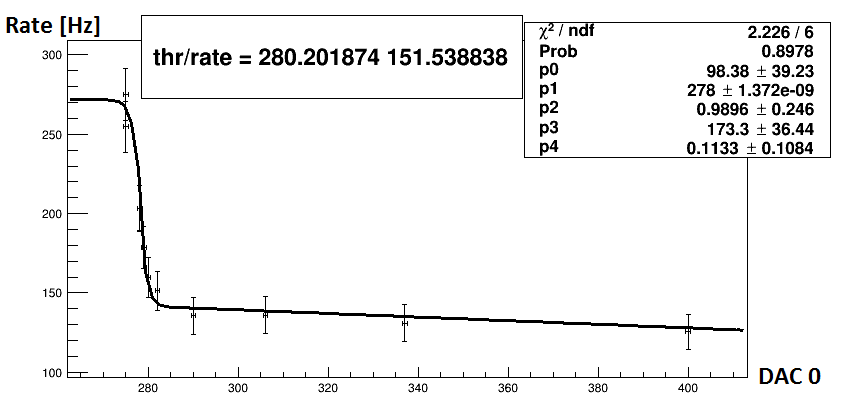} 
\caption{Trigger rate as a function of threshold (DAC0 in MAROC3) for a representative channel and fitting curve as described in text}\label{ingrandimento}
\end{figure}

The single channel trigger rate strongly depends on the discriminator threshold (DAC0 register in MAROC3), at a given channel gain: the rate is several hundreds of Hertz (electronics saturation) for low threshold values, it suddenly decreases (Fig. \ref{ingrandimento}) near the background noise level and then it keeps decreasing roughly linearly. The linear behavior for high threshold values is related (in absence of radioactive sources) to low energy cosmics and thermionic emission (about 1 kHz/PMT with large channel fluctuation).
The threshold dependence of the single trigger rate at fixed channel and gain can be reproduced pretty well be the following Fermi-Dirac distribution summed to a linear term:
\begin{equation} \label{eq:fermi}
f(t)=\frac{a}{1+e^{(t-t_m)/k}}-c \, t \, \theta(t-t_m) + d
\end{equation}
where $\theta()$ is the Heaviside function, $t$ is the threshold variable, $a$, $c$, $d$, $k$ and $t_m$ are parameters; $k$ and $t_m$ correspond to the width of the drop region and the point of sharpest noise drop respectively.

Equation \ref{eq:fermi} has been used to estimate the optimal gains (and their upper limits) for a given threshold in single channel acquisition mode; this strongly reduce the $t-g$ phase space; the search algorithm is detailed in appendix \ref{app:binary}. Thr procedure requires a couple of hours for 1024 channels.

\subsubsection{Trigger rate equalization}

Once the $t-g$ phase space is constrained, a second step equalizes the rate of each channel contributing
to the trigger formation when electronics run in multihit trigger mode: an adaptive gain and threshold scan
is adopted to converge quickly toward the optimal uniformity response to a flood irradiation, as detailed in appendix \ref{app:equal}.

\begin{figure}[htb]
  \centering\includegraphics[scale=0.35]{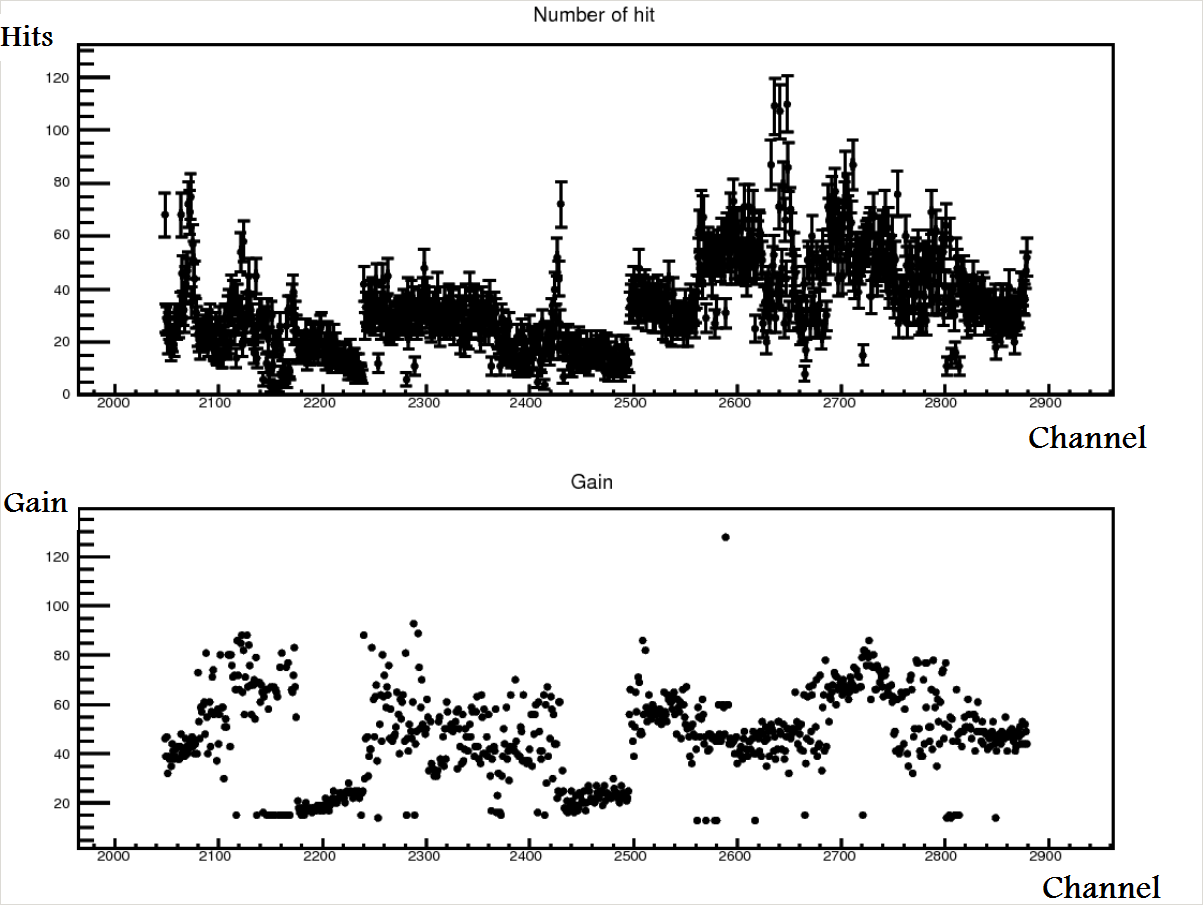} 
\caption{Trigger rate for each channel (top plot) and corresponding set gains (bottom plot), at the beginning of the gain scan iterative procedure.}\label{init_rate}
\end{figure}

\begin{figure}[hbt]
\centering\includegraphics[scale=0.9]{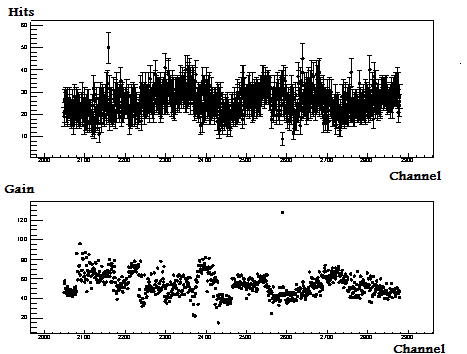} 
\caption{Trigger rate for each channel (top plot) and corresponding set gains (bottom plot), at the end of the gain scan iterative procedure}\label{rate}
\end{figure}

Once this procedure has been completed, a single channel gain scan has been executed, without source.
After a loop of acquisitions, of 10 s each, gain maps are updated again in order to obtain a configuration
in which the contribution to the trigger, of each channel, is the same.\\
In Fig. \ref{init_rate} and Fig. \ref{rate} are reported respectively the channel rate (Number of hits) in the initial and final configuration, for each channel.

\subsection{Local energy Calibration}\label{sec:locenergy}

\begin{figure}[htb]
\centering
\includegraphics[width=13cm]{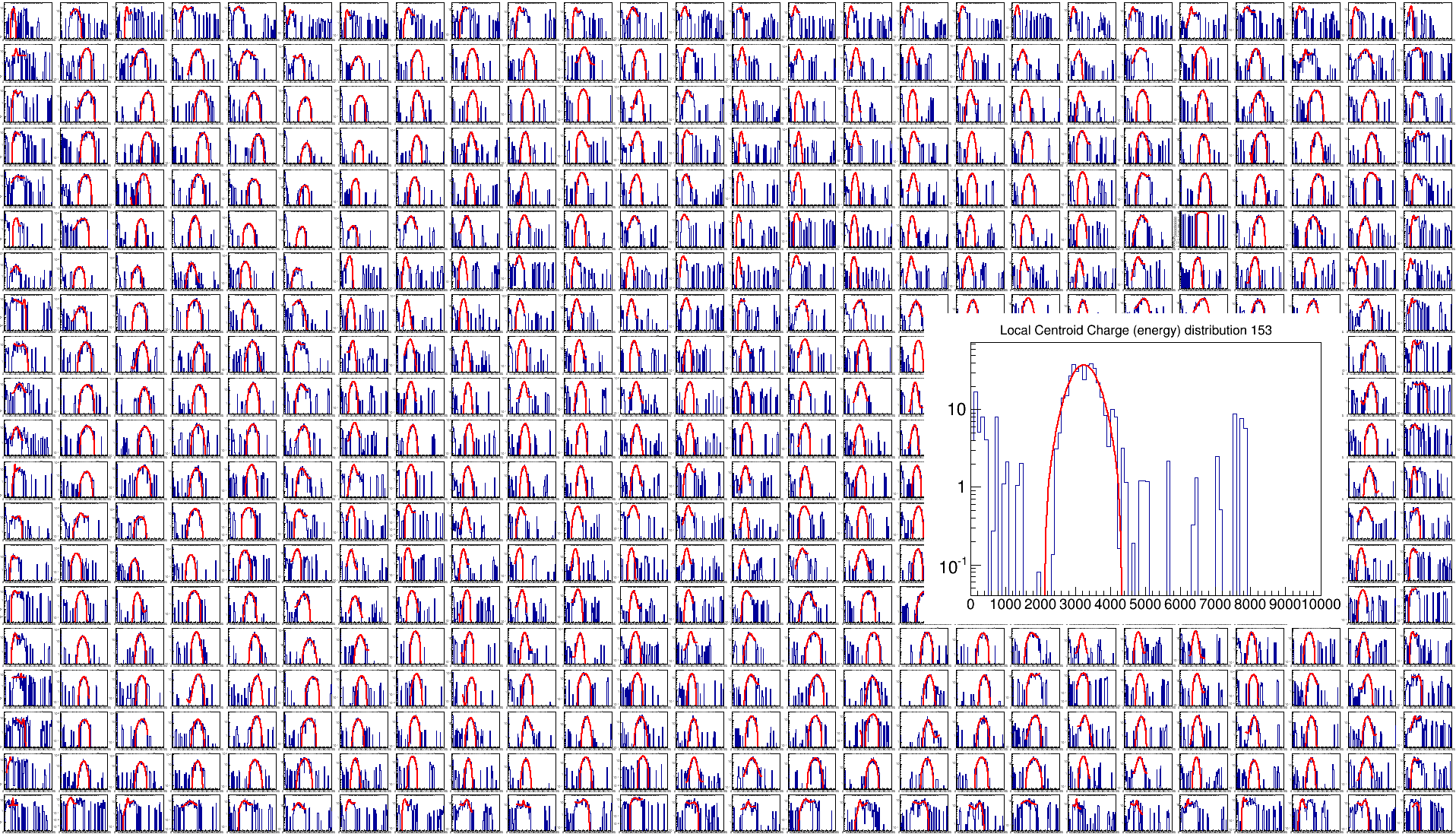} 
\caption{$\ce{^{133}Ba}$ centroid charge distribution, background subtracted, in each local region defined in Big Head. The red curves represents the gaussian fit of the photopeak. The distribution of a single sector is zoomed to show details.}\label{ba_subtracted}
\end{figure}

\begin{figure}[htb]
\centering
\includegraphics[scale=0.25]{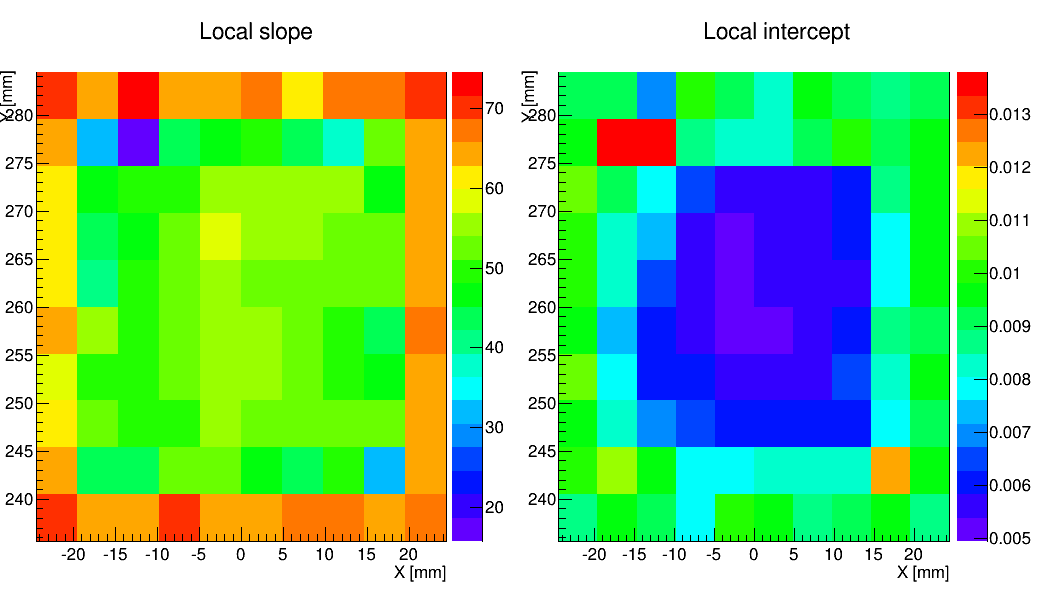} 
\caption{Slopes and intercepts of the local calibration lines for each region of the Small Head detector}\label{local_fit_d1}
\end{figure}

When single channel charges are equalized, they can be summed up to obtain the charge avalanche of the single photoelectic event
which is expected to be proportional (or strongly correlated) to the gamma photon energy. However, intrinsic dis-uniformity due mainly
to detector segmentation requires a local energy calibration, which depends on the position of the photoelectric event;
The procedure for such calibration is summarized by the following main steps:
\begin{enumerate}
\item three (or more) different monochromatic sources
  ($\ce{^{57}Co}$, $\ce{^{133}Ba}$ and $\ce{^{99m}Tc}$), which provide three points for the calibration,
  are acquired in flood-like irradiation configuration, without the detector collimator.
  \item A background run is also acquired. 
\item For each of these runs the number of events is chosen to achieve a good statistics
  which permits the appropriate virtual detector segmentation: the detector surface is
  divided in virtual regions that roughly correspond to the PMT anodes:
  a trade off between energy positional variability and time affordable
  statistics.
\item The centroid and total charge (energy) of each photoelectric event (corrected by the border effects as
  discussed in appendix \ref{app:border}) are computed and then
  the energy distribution of the photoelectric event is produced for each virtual region
  according to the centroid position.
  \item The background energy distribution is subtracted (properly normalized) to the single source distributions, of each virtual region; example is shown in Fig. \ref{ba_subtracted}.
  \item A gaussian fit is performed on the photopeak of the energy distribution in each region (for each background subtracted, source run),
    both for the Big and Small Head.
  \item The mean charge of the photopeaks corresponds to a known energy value, so the parameters, $a$ and $b$, of the linear calibration function $E=a+bC$ can be estimated, by trivial algebra, for each independent region (see Fig. \ref{local_fit_d1} for the Small Head, the Big Head result is very similar).

\end{enumerate}

\begin{figure}[htb]
\centering
\includegraphics[width=12.5cm]{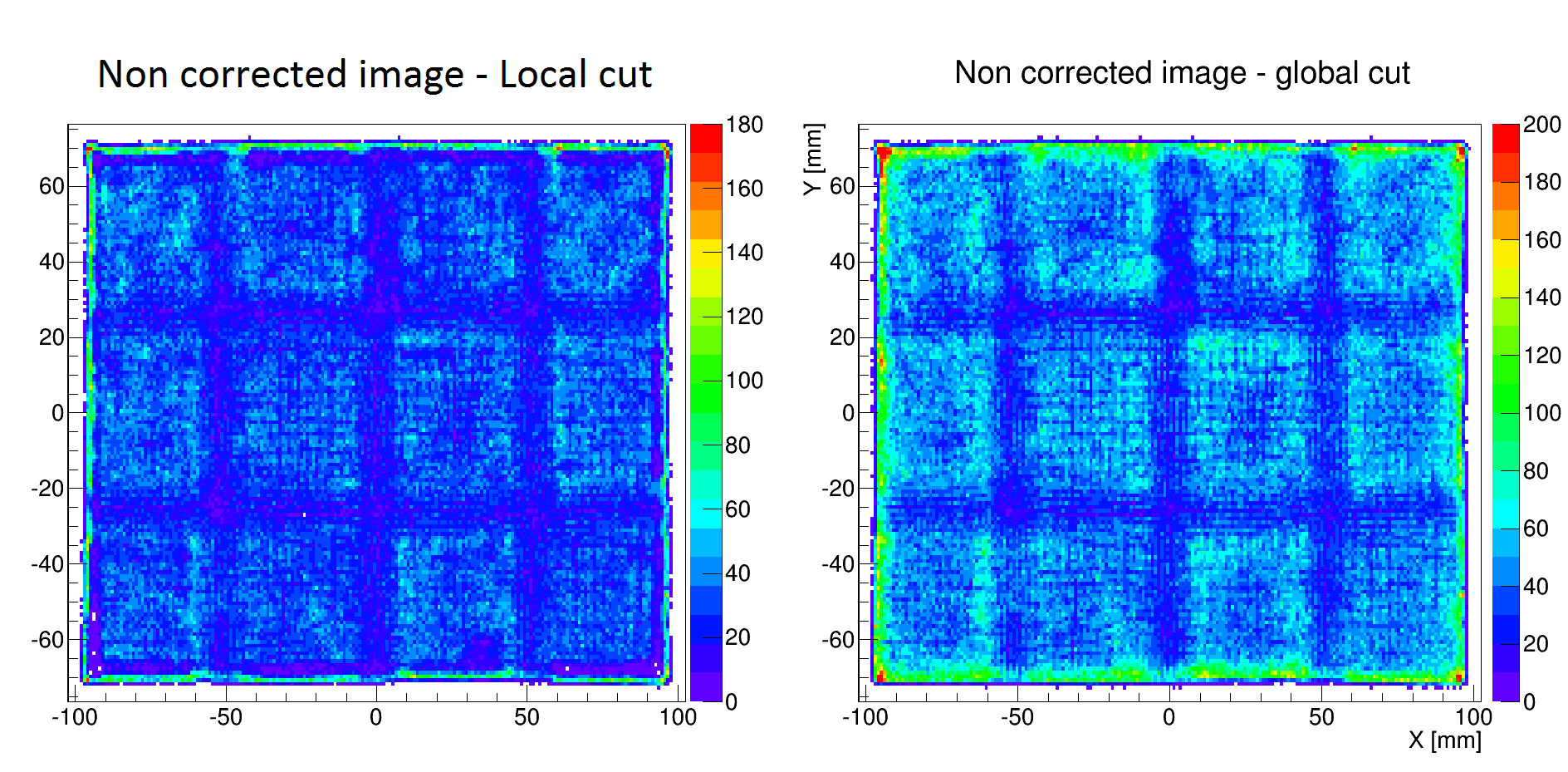} 
\caption{Image obtained from a $\ce{^{57}Co}$ flood, with and without local energy cuts, Big Head}
\label{colocal}
\end{figure}

The above linear calibration function is used to better selects the photoelectric events which sits below the energy
corrected photopeak.
Fig. \ref{colocal} report the image obtained from a $\ce{^{57}Co}$ flood, with and without local energy calibration:
the artifacts of the cracks between PMTs are significantly reduced, with the energy calibration.

\subsection{Centroid Correction}\label{sec:cencorr}
\begin{figure}[H]
\centering
\includegraphics[width=12.5cm]{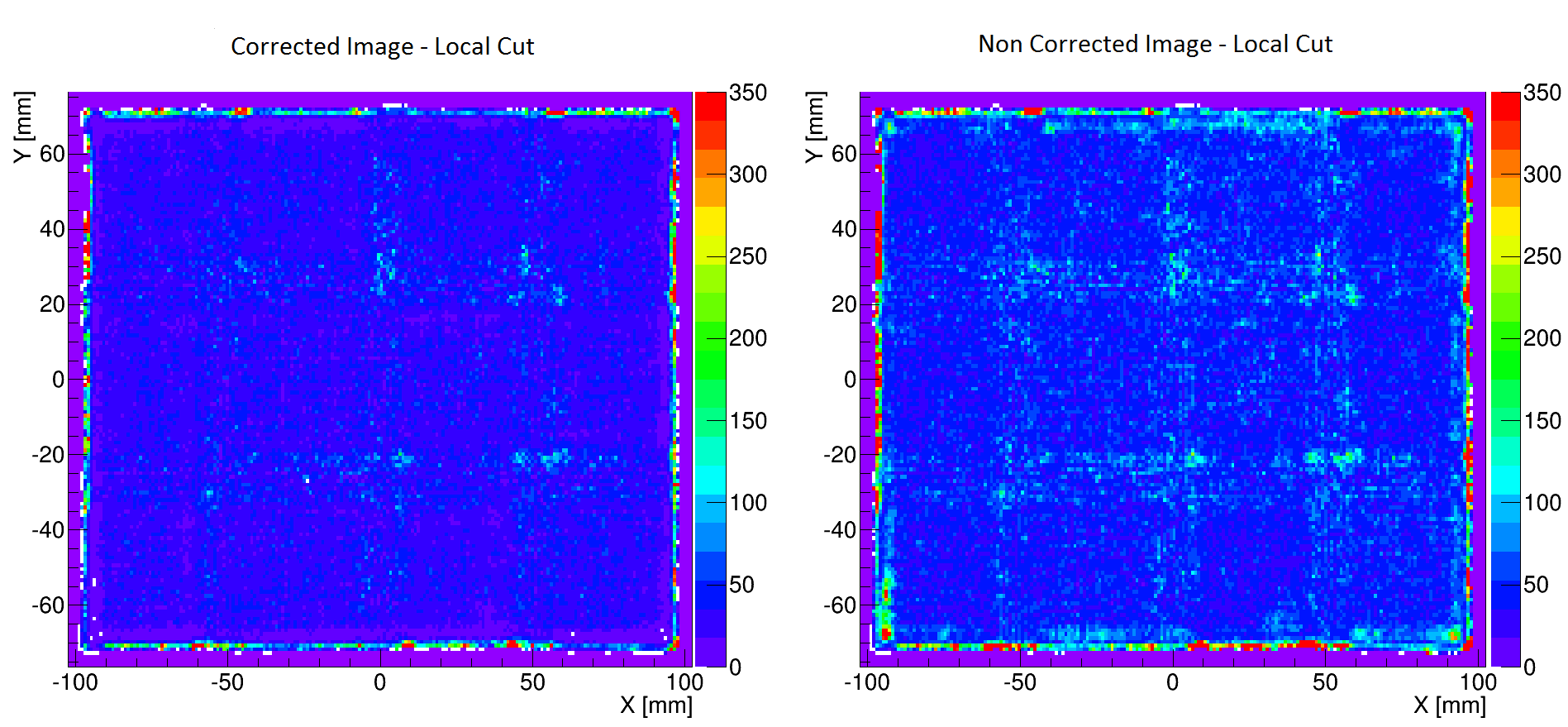} 
\caption{Image of $\ce{^{57}Co}$ flood with and without centroid corrections, Big Head.}\label{centroidcorr}
\end{figure}

The image obtained from a flood, is not yet uniform: the fluctuations of the number of centroids in each pixel
are higher then expected from statistical fluctuations.
In order to reduce the residual systematic non-uniformity of the reconstructed images,
multiplying factors $F_i$ are estimated for each pixel i:
\begin{equation}
F_i=\begin{cases} \frac{\sum_{i=1}^P N_i/P}{N_i}, & \mbox{if }N_i\ne0 \\ 0, & \mbox{if }N_i=0
\end{cases}
\end{equation}
where $N_i$ is the content of the pixel i of the image obtained with a uniform $\ce{^{99m}Tc}$ flood and $P$ is the total number of pixels.
Fig. \ref{centroidcorr} shows the image of $\ce{^{57}Co}$ flood with and without centroid corrections.

\section{Summary and conclusions}
The wide availability and the extensive use of screening mammography have resulted in an earlier diagnosis of breast cancer and in a longer life expectancy. Despite technical improvements and major advantages associated with the use of mammography and other radiological techniques, this technologies have some limitations in clinical practice, especially in women with dense breast tissue, implants, severe dysplastic disease, or significant architectural distortion following breast surgery or radiation therapy. Different noninvasive imaging techniques have been developed to overcome these limitations \cite{bib:tc99}.\\
Molecular Breast Imaging showed to have significantly improved sensitivity with respect to mammography while maintaining the same specificity in screening for women with dense breast \cite{rhodes}.\\
A new asymmetric dual detector system has been implemented in the past years to improve the diagnostic performances: the detectors layout allows performing better spot compression (without additional patient pain), getting one detector closer to the lesion with significant increase in the efficiency and consequently the SNR \cite{bib:garibaldi}.\\
The imaging system is based on segmented sensors and independent channel readout electronics.
The additional flexibility offered by the electronics may result in high quality image processing and formation,
when combined to ad-hoc, dedicated and optimized calibration procedures at the hardware and software level.

In operating conditions these calibration procedures are supposed to be performed periodically and therefore their duration
need to be minimized to make them clinically affordable.

The above described approach tries to take all these aspects into account:
the optimal electronics operation point (in terms of thresholds and gains) is essentially chosen to maximize the uniformity of
single channel responses when the detector is irradiated by a uniform field, minimizing signal inefficiency.
Image formation is further improved by locally calibrated energy distribution of the photoelectric events
and residual spatial equalization of the single image pixels.

The above described procedures will be adopted on the next foreseen upgrade of the detector that will
move to SiPM optical photo sensors, coupled to the same acquisition electronics, to obtain a more compact
sensor at reduced cost. The updated device will be integrated into a tomosynthesis system to get almost
simultaneously morphological and functional information for an ultimate
multimodality imaging, for superior breast cancer diagnosis.


 \appendix

\section{Binary scan and gain-threshold relation}
\label{app:binary}

For each channel at a given gain, a scan in DAC0 has been executed in two experimental conditions:
\begin{itemize}
\item  {\bf flood irradiation}: in laboratory, a sealed point-like source of $\ce{^{57}Co}$, with approximately 300 kBq activity  at 1 m from the detector without collimator has been used; in clinics, a source of $\ce{^{99m}Tc}$ of similar activity in a small vias at similar distance has adopted.
\item  {\bf background irradiation}: same conditions, without source, similar to the pedestal run, but with internal self-triggering.
\end{itemize}

To speed up the scan without losing relevant data (concentrated in the sharp rate drop region as shown in fig. \ref{ingrandimento}),
a binary scan approach has been adopted:
\begin{enumerate}
\item the first two rates $R_L$ and $R_H$ are estimated at low and high thresholds $t_L$ and $t_H$ respectively; $R_L$ and $R_H$ correspond to very noisy and ``noise free'' conditions;
\item \label{en:tn} then the next threshold $t_N$ is computed by $t_N = (t_H + t_L)/2$. The rate $R_N$ measured at $t_N$ is compared to the previous ones;
\item if $(R_L - R_N) < (R_N - R_H)$ then $R_L$ and $t_L$ are set to $R_N$ and $t_N$ respectively; if the inequality is false, $R_H = R_N$ and $t_H = t_N$;
\item step \ref{en:tn} and successive are iterated until the new $t_N$ is equal to the previous one within 2 threshold units.
\end{enumerate}
This procedure provides a quick, $\sim 10$ point scan, accumulating in the most variable rate region, as shown in fig. \ref{ingrandimento}.

The measured rate distributions are fitted by function \ref{eq:fermi}; the fitting functions, for a given channel and gain, with and without radioactive source are subtracted; the DAC0 corresponding to the maximum of the difference (larger than both estimated $t_m$'s) is considered the optimal threshold $t_O$ for that gain and channel, while $t_m$ is the minimum admissible threshold.

\begin{figure}[htb]
\centering\includegraphics[scale=0.25]{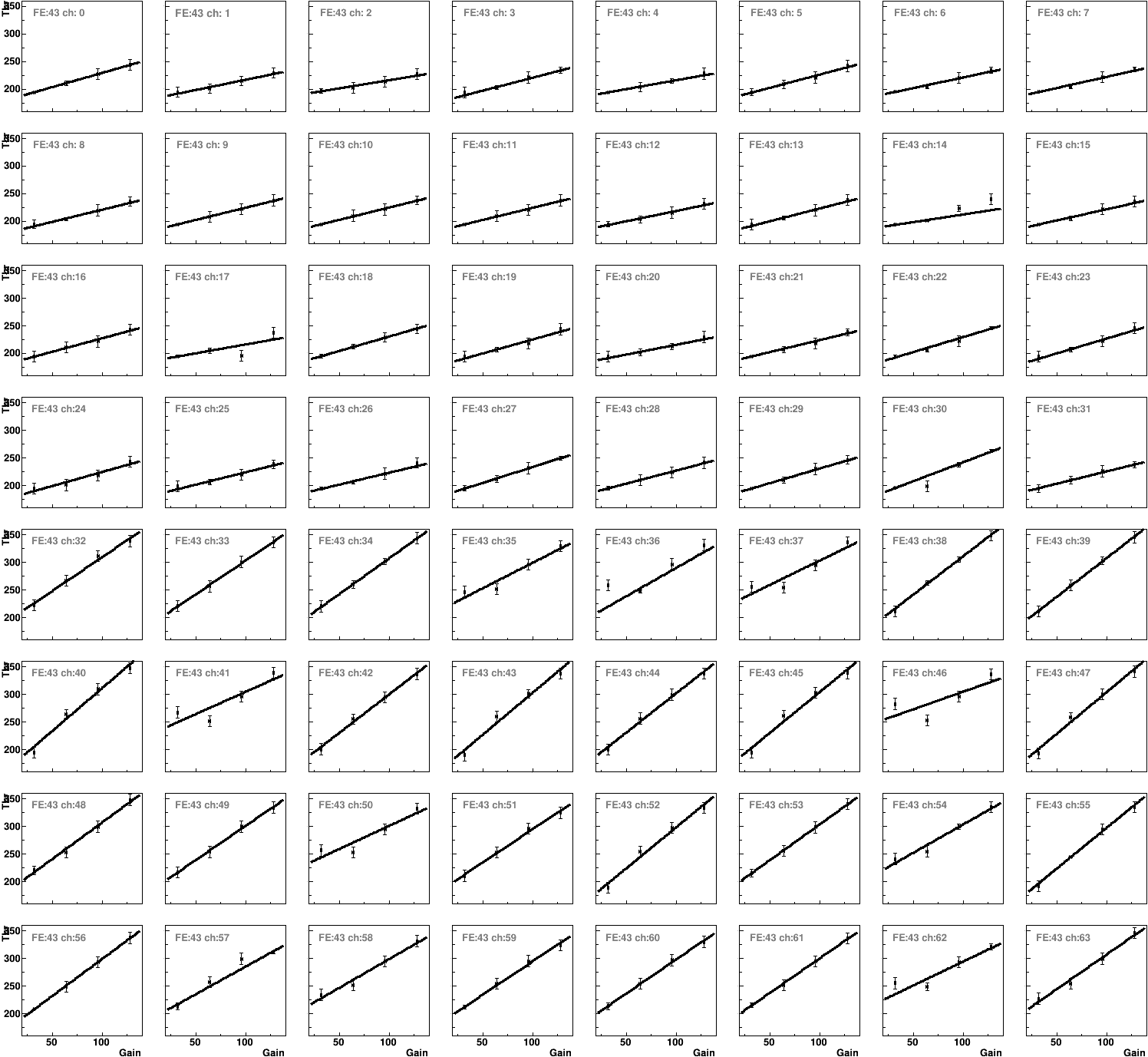} 
\caption{Optimal threshold $t_O$ versus gain $g$ for representative channels; points are estimated values, as described in the text, while the line represents the fit.}\label{fig:thr-vs-gain} 
\end{figure}

The procedure is repeated for at least 3 different gains in the range of $0.5 - 4$ uniformly distributed,
obtaining the optimal threshold $t_O(g,i)$ depending on gain $g$ and channel $i$; $t_O$ (as well as $t_m$) is linear in
gain\footnote{At high gain the loss of linearity in the response is due to saturation effects.} as shown in fig. \ref{fig:thr-vs-gain} and therefore it can be reliably interpolated in $g$.

Unfortunately, groups of 64 channels must share the same DAC0 and therefore different optimal thresholds
and gains cannot be applied simultaneously. However, $t_O(g,i)$ is inverted to get the optimal 
gain $g_O(t,i)$ that can be used at a given threshold; similarly for $t_m$ and corresponding maximum gain $g_M$: beyond it, the electronic noise start to becomes relevant.

\section{Dynamic gain equalization procedure}
\label{app:equal}

Once the optimal and maximum gains of each channel have been estimated as a function of the threshold (appendix \ref{app:binary}),
the following adaptive procedure is used to get a uniform single channel rate response to uniform irradiation conditions (either flood or background), in normal acquisition multihit trigger mode, which tends to select cosmic and environmental radioactivity
events but also random multiple coincidences from the single anode dark rate.

The uniformity is obtained acting on the single channel gains, and the thresholds (one for groups of 64 channels).
The main steps are outlined here:
\begin{enumerate}
\item Assuming all channel at unitary gain, the average optimal thresholds is set on each readout cards\footnote{Channels with pathological  gain-threshold curves are eventually masked.}.
\item Set the optimal single channel gain corresponding to the chosen threshold.
\item\label{gain:runbck} Run a short (few thousand events) background acquisition
\item Extract the single channel trigger rate distribution $R_c$ from the data, which depends from the channel gain and threshold.
\item Estimate the average trigger rate $R_L$ of a statistically significant fraction ($\le 10$\%) of the channels with
  the lowest trigger rates.
\item Change the gain of each channel, proportionally to the difference $R_L - R_c$, avoiding to exceed the fiducial gain range between a fraction above zero and below a fraction from the maximum admissible gain at the running threshold\footnote{The lowest fraction is assumed at a gain of 0.2 and the upper fraction is 0.1 from the admissible maximum.}.
\item Set the new gains and the average optimal threshold on each readout cards.
\item Iterate from step \ref{gain:runbck} above, until the single channel trigger rate RMS does change below a desired value\footnote{RMS/MEAN change below 1\% between iterations}.
\end{enumerate}

The above procedure generally converge in few hours, if the acquisition conditions are stable.

\section{Border Correction}\label{sec:bord_corr}
\label{app:border}

\begin{figure}[H]
\centering
\includegraphics[width=12cm]{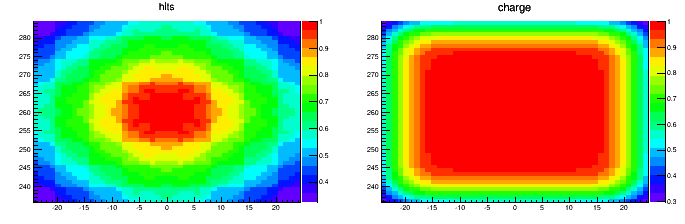} 
\caption{Left: fraction of anodes involved in every hits, normalized to the anodes number. Right: charge correction weight $w$, for the Small Head, as function of the position}\label{bord}
\end{figure}

In a scintillation event the spatial charge distribution involves several pixels so, if the event has happened near the borders of the detector, part of the charge released in that event will not be collected. If the event belongs to the central part of the detector, from the charge distribution the centroid can be reconstructed properly but if the photon hits on the borders of the detectable surface, the reconstructed photon energy can be wrong, because of this missing charge.\\
This effect can be mitigated partially by introducing a position dependent correction to the charge released in each event.
Centering a 2D gaussian in each anode position, a weighting factor $w$ can be computed for each pixel whose distance is less then R (radius used for the centroid calculation) from the anode:
\begin{equation}
w=\frac{1}{2\pi\,\sigma^2}\cdot e^{-\frac{(x-x_{pix})^2-(y-y_{pix})^2}{2\,\sigma^2}}
\end{equation}
where $(x,y)$ and $(x_{pix},y_{pix})$ are respectively the anode and pixel positions on the detector. \\
The correction to the number of hits is simply given, for each pixel, by:
\begin{equation}
h=\frac{n_a}{N_a}
\end{equation}
where $n_a$ is the number of anodes whose distance is less then R and $N_a$ is the total number of anodes of the detector.\\
This border effect is dominant for the Small Head, because of its dimension; in Fig. \ref{bord} are shown the hits number and charge correction, for each pixel.
It is a correction to the border effect, in fact, while in almost the whole detector the correction factor is unitary, close to the edges it gets smaller.



\begin{thebibliography}{99} 

\bibitem{globascan} F. Bray et al.,
  \textbf{Global cancer statistics 2018: GLOBOCAN estimates of incidence and mortality worldwide for 36 cancers in 185 countries},
  CA: A Cancer Journal for Chinicians, 12 Set. 2018 (doi:10.3322/caac.21492)

\bibitem{bib:OConnor2} M. K. O'Connor, C. B. Hruska
  \textbf{Nuclear imaging of the breast: Translating achievements in instrumentation into clinical use},
  Medical Physics, Vol. 40, No. 5, May 2013
  

\bibitem{bib:oconn15} M. K. O'Connor,
  \textbf{Molecular breast imaging: an emerging modality for breast cancer screening},
  Breast Cancer Management 4(1), pg. 33-40, 2015

\bibitem{bib:freer} P. E. Freer,
  \textbf{Mammographic Breast Density: Impact on Breast Cancer Risk and Implications for Screening},
  RadioGraphics, vol. 32, n. 2 (2015) pg. 302-315 (doi:10.1148/rg.352140106)

\bibitem{bib:vourtsis18}
  A. Vourtsis, W. A. Berg,
  \textbf{Breast density implications and supplemental screening},
  European Society of Radiology, 2018 Sep 25 (doi: 10.1007/s00330-018-5668-8)

    \bibitem{bib:costs} A. Vlahiotis et al.,
  \textbf{Analysis of utilization patterns and associated costs of the breast imaging and diagnostic procedures after screening mammography}, ClinicoEconomics and Outcomes Research, vol. 10, pg. 157-167, March 2018 (doi:10.2147/CEOR.S150260)
  
\bibitem{bib:Kim11} B. S. Kim
\textbf{Usefulness of breast-specific gamma imaging as an adjunct modality in breast cancer patients with dense breast: a comparative study with MRI.} Ann Nucl Med. 2011 Oct 18.

\bibitem{Khalkhali} I. Khalkhali et al.
  \textbf{Prone scintimammography in patients with suspicion of carcinoma of the breast} J Am Coll Surg, vol. 178, pp. 491, 1994

  \bibitem{scop}F. Scopinaro et al.
\textbf{Imaging probe for breast cancer localization}, Nucl Instr Meth A., vol. 497, pp. 114-121, 2003

\bibitem{devinc}G. De Vincentis et al.
\textbf{Results of clinical trials with SPEM}, Nucl Instr Meth A, vol. 497, pp. 46-50, 2003


\bibitem{bib:tc05}R. Teillefer,
  \textbf{Clinical Applications of $^{99m}$Tc-Sestamibi Scintimammography}, Seminars in Nuclear Medicine, 2005


\bibitem{bib:garibaldi}F. Garibaldi et al.
  \textbf{A novel high resolution and high efficiency dual head detector for molecular breast imaging: New results from clinical trials},
  Nuclear Instruments and Methods in Physics Research A 617 (2010) 227 -229

\bibitem{bib:pmt} Hamamatsu,
  \textbf{Flat Panel Type Multianode PMT Assembly, H8500 SERIES/H10966 SERIES}, 2011, available on the HAMAMATSU web page (Nov/2018): https://www.hamamatsu.com/
  (file name H8500\_H10966\_TPMH1327E.pdf)

\bibitem{bib:argentieri10} A. G. Argentieri et al.,
  \textbf{A Multichannel Compact Readout System for Single Photon Detection: Design and Performances} Nucl. Instr. Meth. Phys. Res., 2010

\bibitem{bib:altera} Altera,
  \textbf{Cyclone II Device Handbook, Volume 1}, 2008, available on the INTEL web page (Nov/2018): https://www.intel.com/
  (file name cyc2\_cii5v1.pdf)
  
\bibitem{bib:maroc} Omega,
  \textbf{MAROC3 Datasheet}, web site (Nov/2018): http://omega.in2p3.fr/

\bibitem{bib:barillon}Barrillon et al.
  \textbf{MAROC: Multi-Anode ReadOut Chip for MaPMTs}, Nuclear Science Symposium Conference Record, 2006, IEEE, vol. 2, pag. 809-814

\bibitem{bib:root}Rene Brun and Fons Rademakers,
  \textbf{ROOT - An Object Oriented Data Analysis Framework},
  Proceedings AIHENP'96 Workshop, Lausanne, Sep. 1996, Nucl. Inst. \& Meth. in Phys. Res. A 389 (1997) 81-86. See also http://root.cern.ch/

\bibitem{bib:iaea}International Atomic Energy Agency
  \textbf{Quality assurance for Spect Systems}, Vienna, 2009

\bibitem{pani} R. Pani et al, \textbf{Factors Affecting Hamamatsu H8500 Flat Panel
  PMT Calibration for Gamma Ray Imaging}, IEEE TRANSACTIONS ON NUCLEAR SCIENCE, Vol. 54, Number 3, Pg. 438, June 2007

  \bibitem{bib:tc99}R. Teillefer.
  \textbf{The role of $^{99m}$Tc-sestamibi and other conventional radiopharmaceuticals in breast cancer diagnosis} Seminars in Nuclear Medicine, Volume 29, Issue 1, January 1999, Pages 16-40

\bibitem{rhodes} D.J. Rhodes et al.
  \textbf{Dedicated Dual-Head Gamma Imaging for Breast Cancer Screening in Wmen with Mammographically Dense Breast}, Radiology, Vol. 258, Number 1, 2011

\end{thebibliography}
\end{document}